\documentclass[11 pt]{article}
\usepackage{latexsym}
\usepackage{epsfig}
\usepackage{amscd}
\usepackage{amssymb}

\newcommand{\ws}{\hat s}
\newcommand{\tM}{\hat M}
\newcommand{\tg}{\hat g}
\newcommand{\pd}{\partial} 
\newcommand{\re}{{\mathbb R}}
\newcommand{\scri}{{\mathcal I}}

\newcommand{\cN}{{\mathcal N}}
\newcommand{\cO}{{\mathcal O}}
\newcommand{\cU}{{\mathcal U}}
\newcommand{\tog}{\tg^\circ}

\newcommand{\tograd}{\hat \nabla^\circ}
\newcommand{\tq}{{\hat q}} 
\newcommand{\toq}{\tq^\circ}
\newcommand{\cC}{{\mathcal C}}
\newcommand{\hC}{\hat {\rm C}}

\newcommand{\mw}{{\langle \mu \rangle_{t_0}}}
\newcommand{\om}{\Omega_{\mathbf m}}
\usepackage{verbatim}
\makeatletter
\renewcommand{\section}{\@startsection{section}{1}{0mm}%
{-0.5\baselineskip}{1 pt}{\normalfont\large\bfseries}}
\renewcommand{\subsection}{\@startsection{subsection}{2}{0mm}%
{-0.1\baselineskip}{1 pt}{\normalfont\bfseries}}
\renewcommand{\subsubsection}{\@startsection{subsubsection}{3}{0mm}%
{-0.1\baselineskip}{1 pt}{\normalfont\bfseries}}
\renewcommand{\paragraph}{\@startsection{paragraph}{4}{0mm}%
{-0.1\baselineskip}{1 pt}{\normalfont\itshape}}
\renewcommand{\subparagraph}{\@startsection{subparagraph}{5}{0mm}%
{-0.1\baselineskip}{1 pt}{\normalfont}}
\title{{Positive Mass from Holographic Causality} 
\thanks{Alberta-Thy-08-02, hep-th/0204198}}
\author{Don N. Page${}^{a}$\footnote{Fellow, Canadian Institute for Advanced  
Research} \footnote{email: don@phys.ualberta.ca} \ , 
Sumati \ Surya${}^{\, a, b}$\footnote{e-mail: ssurya@phys.ualberta.math.ca}
\ and  Eric \ Woolgar${}^{\, b, a}$\footnote{e-mail: ewoolgar@math.ualberta.ca}
\\ 
\\
${}^a$ Theoretical Physics Institute, University of Alberta\\ Edmonton,
AB, Canada T6G 2G1  
\\
\\
${}^{b}$
Dept.\ of Mathematical Sciences, University of Alberta\\
Edmonton, AB, Canada T6G 2G1}
\date{}

\begin{document}

\maketitle
\begin{abstract}
For $n+1$ dimensional asymptotically AdS spacetimes which have
holographic duals on their $n$ dimensional conformal boundaries, we
show that the imposition of causality on the boundary theory is
sufficient to prove positivity of mass for the spacetime when $n \geq
3$, without the assumption of any local energy condition. We make
crucial use of a generalization of the time-delay formula calculated
in gr-qc/9404019, which relates the time delay of a bulk null curve
with respect to a boundary null geodesic to the Ashtekar-Magnon mass
of the spacetime.  We also discuss holographic causality for the negative
mass AdS soliton and its implications for the positive energy
conjecture of Horowitz and Myers.
\end{abstract}
\vspace{0.1 cm}

Positive energy theorems play a central role in gravitational
theories, since they determine the classical stability of physical
spacetimes within given asymptotic classes.  Indeed, alternative
versions of the proof of positivity exist, using surprisingly distinct
techniques. In the asymptotically flat case, among the many
approaches, one has the
proof of positivity using minimal surfaces, due to Schoen and Yau
\cite{sy}, Witten's supersymmetry-inspired proof using spinorial
techniques \cite{witten}, Geroch's approach using the inverse mean
curvature flow \cite{geroch,hi} and Penrose, Sorkin and Woolgar's
approach using the focusing of null geodesics near conformal infinity
$\scri$ \cite{psw}. Similarly, in the asymptotically AdS case one has
the spinorial proof due to Abbott and Deser \cite{ad}, and a proof
based on null focusing techniques in \cite{eric}.

In this paper, we present a new proof of positivity of energy for
asymptotically AdS spacetimes based on the assumption of holography
\cite{susskind,tHooft,sussbig,bousso}.  While the AdS/CFT
correspondence \cite{mal,gkp,W,agmoo} is a particular realization of
holography, our result does not depend on the specifics of the
holographic correspondence.

In holographic theories, causally related events in the $n+1$
dimensional bulk theory have interpretations in the $n$ dimensional
holographic dual theory and vice versa.  Questions of causality in
holographic theories have been discussed in many papers
\cite{horit,kablif,bakrey,gidd,pst,susstoumb,lmr,giddlipp,kgt}.  For
example, Kleban, McGreevy and Thomas \cite{kgt} propose a relationship
between causality in the holographic dual theory and matter energy
conditions in the bulk.

	We show that the requirement of causality in the holographic
dual theory is in fact sufficient to prove that 
a certain weighted time-average of the mass 
of the spacetime is non-negative, given the assumption of a holographic
correspondence, but without the assumption of any energy conditions. 
In the special case when the matter flux on the conformal boundary
$\scri$ vanishes, this means that the instantaneous mass itself is
non-negative.
Our proof is based on a generalization to arbitrary
spacetime dimensions $n+1 \geq 4$ of an expression for time delay in
\cite{eric}, which relates the time difference between a null geodesic
on the conformal boundary $\scri$ and a nearby bulk null curve to the
Ashtekar-Magnon mass.

As in the AdS/CFT correspondence, we shall take a holographic
correspondence to imply that certain fields in the (conformally
completed) bulk theory, evaluated at a point on the conformal
boundary, correspond to operators in the dual, or ``boundary'', theory
localized at the same point.  We further assume {\it holographic
causality}: if two points on the conformal boundary are causally
related via the bulk causal structure, then there is a causal relation
(e.g. non-commutation) between the two corresponding operators of the
dual boundary theory. The boundary theory is thus defined to be causal
only if these two operators are localized at points that are causally
related in the conformal boundary causal structure.  More crudely,
causality implies that a signal cannot go through the bulk faster than
it can go along the conformal boundary.

We begin by considering the pair: a point S on $\scri$, dubbed the
source, and an observer world-line $R$ on $\scri$, dubbed the
receiver.  Let $\Delta t$ be the time taken by a fastest null geodesic
$\gamma$ to go from S to R along $\scri$ and $\Delta t'$ the time
taken by a bulk null curve $\gamma'$ to go from S to R.  If $\Delta t
> \Delta t'$, this means that R will receive a signal from S via the
bulk along $\gamma'$ before it receives the fastest boundary signal
$\gamma$, thus violating holographic causality. Holographic causality
therefore requires that the time delay $\Delta T= \Delta t'-\Delta t$
between the fastest boundary null geodesic and any bulk null curve 
must always be non-negative \cite{kgt}.

When the spacetime is pure AdS, the time delay between the fastest
boundary signal and any bulk signal is non-negative. It vanishes when
the bulk signal is along a bulk null geodesic, but this can only join
antipodal points on the boundary. Since the strictest version of
AdS/CFT holography holds for pure AdS bulk spacetimes (in 3, 5 and 7
spacetime dimensions), holographic causality is satisfied.  However,
one must also be able to interpret, in the boundary theory, bulk
perturbations about the global AdS background
\cite{balasub,horit}. Such perturbations would affect this exact
matching and hence could lead to a violation of holographic causality.
Intuitively one would expect that if global AdS is perturbed by the
addition of positive mass, then causal geodesics in the bulk will
focus and hence be non-maximal, thus maintaining holographic
causality.  In fact what we will find is the converse of this
intuition --- namely, the fact that bulk causal curves are non-maximal
implies positivity of mass.
\begin{figure}[ht]
 \centering
\resizebox{1.5in}{!}{\includegraphics{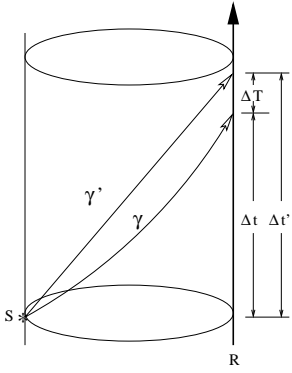}}
\vspace{0.5cm}
\caption{{\small In order to preserve holographic causality, the time
delay $\Delta T=\Delta t'-\Delta t$ must be
non-negative.}\label{timedelay.fig}}
\end{figure}

To make our treatment self contained, we start by reminding the reader
of the conformal  definition of an asymptotically AdS spacetime in $n+1$
spacetime dimensions for $n+1 \geq 4$ \cite{am,ashdas}; we use the
definition in \cite{ashdas} rather than the weaker one in \cite{am}.   

\noindent {\bf Definition:}\label{ads} An $n+1$ dimensional spacetime
$(M, g_{ab})$ is said to be asymptotically AdS if there exists a
manifold $\tM$ with conformal boundary $\scri$ , with a (sufficiently)
smooth metric $\tg$ and a diffeomorphism from $M$ to $\tM - {\scri}$
such that \renewcommand{\labelenumi}{(\theenumi)}
\renewcommand{\theenumi}{\roman{enumi}}
\begin{enumerate}
\item There exists a function $\Omega$ on $\tM$ such that
$\tg_{ab}=\Omega^2g_{ab}$ on $M$.
\item $\scri=\pd \tM$ has topology $S^{n-1} \times \re$, $\Omega=0$ on
$\scri$ and $\nabla_a \Omega$ is nowhere vanishing on $\scri$.
\item  $g_{ab}$ satisfies $R_{ab} - \frac{1}{2}Rg_{ab} + \Lambda
g_{ab}=8 \pi T_{ab}$ where $\Lambda <0$  and $\Omega^{1-n}\,T_{ab}$ 
admits a smooth limit to $\scri$, 
\item The Weyl tensor of $g_{ab}$ is such that $\Omega^{3-n} \hC_{abcd}$
is smooth on $\tM$ and vanishes at $\scri$. 
\item For $n=3$ only (where it does not follow from the other
conditions) we additionally require that the conformal isometry group
of $\scri$ is the AdS group.
\end{enumerate}

We confine our attention to the conformal metric $\tg_{ab}=\Omega^2
g_{ab}$ in a small neighbourhood $\cN = \re \times S^{n-1} \times
[0,\epsilon] = \re \times \cU $ of $\scri$, where $\Omega\in
[0,\epsilon]$.  It is convenient to express $\tg_{ab}$ to order
$\cO(\Omega^{n+1})$ in a coordinate frame $x = (\Omega, \vec x)$
adapted to the level surfaces of $\Omega$, where $\vec x= (t, \vec
\omega_{n-1})$ are the coordinates in constant $\Omega$ hypersurfaces
and $\vec \omega_{n-1}$ are the coordinates on the $S^{n-1}$ sphere.
The conformal metric in $\cN$ may be expressed as \cite{hete,fg,HS}
\begin{equation}
\label{falloff}
d{\hat s}^2 = d{\hat s}_{\circ}^2 + \Omega^n \, \, f_{ab}(\vec x) \,
dx^a \, dx^b +{\cal O}(\Omega^{n+1}), 
\end{equation} 
where 
\begin{equation} 
\label{adS}
d{\hat s}_{\circ}^2 = \tog_{ab}\, dx^a dx^b = \Omega^{2}\,\,
ds_{AdS}^2= \frac{1}{1-\Omega^2}\,\,d\Omega^2 -\,dt^2 + (1-\Omega^2)
\,\,d\omega_{n-1}^2,
\end{equation} 
with $ds_{AdS}^2$ the AdS metric, $d\omega_{n-1}^2$ the standard
``round'' metric on an $n-1$ sphere  and $f_{ab}(\vec x)$  a
degenerate symmetric tensor with $f_{a\Omega}=0$. For simplicity  we
use units in which $\Lambda = -\frac{n(n-1)}{2}$.  

As we now show, the trace-free part of $f_{ab}$ is in fact
proportional to the electric part $E_{ab}$ of the Weyl tensor $\hC_{abcd}$, 
\begin{equation} \label{EW}
E_{a b}(\vec x) =\lim_{\rightarrow \scri}\, \, \frac{1}{\Omega^{n-2}}
\, \, {\hC}_{a c b d} \, \, \hat \nabla^c\Omega \, \hat \nabla^d \Omega.   
\end{equation} 
Using the form of the metric (\ref{falloff}), $\hC_{abcd}$  
can be simplified  to
\begin{eqnarray} 
\hC_{abcd}&=&-2\tog_{de} \tograd_{[a} \, \cC_{b]e}^e - \, 
\frac{4}{n-1}\, \, (\tograd_{[d} \, \tog_{c][b}\, \cC_{a]e}^e +
\tograd_{e}\, \tog_{[a|\, [c}\, \cC_{d]\,| b]}^e)     
 \nonumber \\ 
&&
- \, \frac{4}{n(n-1)} \,\,  \tog_{a[c} \, \tog_{d]b}\, \tg^{\circ \,
ef}\, \tograd_{[e}\, \cC_{m]f}^m + \cO(\Omega^{n-1}),     
\end{eqnarray} 
where $\tograd_a$ is the connection compatible with the conformal AdS
metric $\tog_{ab}$, $\hat \nabla_a$ is the connection compatible with
(\ref{falloff}), the tensor $\cC_{ab}^c$ is given by $ \cC_{ab}^c\,
v_c= \tograd_a \, v_b - \hat \nabla_a\, v_b$, and we have used the fact
that AdS has vanishing Weyl curvature, $\hC^\circ_{abcd}=0$.  Using
(\ref{falloff}), (\ref{adS}) and (\ref{EW}), and denoting the induced
metric on $\scri$ by $\toq_{ab}$, we get
\begin{equation}
E_{ab}= -\, \frac{n(n-2)}{2}\,\, f_{ab} + \frac{(n-2)}{2}\,\, \toq_{ab}\, (\tq^{\circ \, cd}\, f_{cd}).
\end{equation} 
Reinserting this into (\ref{falloff}), we obtain
\begin{equation} 
\label{A1}
d{\hat s}^2 = (1+ \Omega^n \, H(\vec x))d{\hat s}_{\circ}^2 - \frac{2 }{n(n-2)}
\, \, \Omega^n\, E_{a b}(\vec x)\,dx^a dx^b - \Omega^n \, H(\vec x)\,
d\Omega^2 +{\cal O}(\Omega^{n+1}), 
\end{equation} 
where $n \, H(\vec x) \equiv \tq^{\circ \, kl}\, f_{kl}$ and where we
have used the fact that $E_{a \Omega}=0$. The definition of
asymptotically AdS spacetimes above, from \cite{ashdas}, 
implies $H(\vec x) =0$, so our proof actually applies to the broader
class of metrics given by (\ref{falloff}).

We now choose S and R so that they project to the antipodes of the
spatial boundary sphere $S^{n-1} \subset \scri$. Next, we use the arc
length $\sigma \in [0,\pi]$ to parameterise the null curves, where
$d\sigma^2 =(1-\Omega^2)^{-1} \,d\Omega^2
+(1-\Omega^2)\,d\omega_{n-1}^2$ is the ``round'' metric on $S^{n}$
with the equator at $\Omega=0$.  Let $\gamma:[0,1] \to \scri$ be one
of the boundary null geodesics  between S and R and let $\gamma': [0,1] \to
\cN$ be a bulk null curve which is a particular variation of
$\gamma$ with $\gamma'(0)=\gamma(0)$. 
Since $d\sigma^2$ is the round metric on $S^{n}$, $\gamma$ projects to
the equator at $\Omega=0$, and we may now choose $\gamma'$ to project
to a nearby geodesic of $d\sigma^2$ in $\cU$. This is always possible
on the round $S^n$, since geodesics with the same starting point will
focus at the antipodal point. Thus, for
the varied curve $\gamma'$,  $\Omega(\sigma) = \om \sin(\sigma)$, 
where $\om$ is the farthest that $\gamma'$ ventures into the bulk. For
a sufficiently small $\om >0$ we then obtain the following expression for
the time delay in $\gamma'$ with respect to $\gamma$:
\begin{equation} 
\label{TD1}
\Delta T = -\frac{\om^{n}}{n(n-2)} \int_0^\pi
\sin^n(\sigma)\,\, E_{ab}(\vec x) \, \gamma^a 
\gamma^b\, \, d\sigma  + \cO(\om^{n+1}), 
\end{equation} 
where $\vec x$ and $\gamma^{a}=\frac{dx^a}{d\sigma}$ are functions of
$\sigma$ along the boundary null geodesic $\gamma$.  
The requirement of holographic causality implies
a non-negative time delay, so that the right hand side of (\ref{TD1})
must be non-negative.  We now show that the integrand is related to
the Ashtekar-Magnon mass.

The Ashtekar-Magnon(AM) mass is defined as \cite{am,ashdas}  
\begin{equation}
\mu[C]:= -\frac{1}{8\pi (n-2)} \int_{C} E_{a b} \chi^a 
\chi^b dS, \label{AM1}
\end{equation}
where $C$ is a space-like cut of $\scri$, $dS$ the associated volume
element and $\chi^a$ a conformal timelike Killing vector on $\scri$.  In the
coordinate frame $\vec x = (t, \vec \omega_{n-1})$ we use above, a
natural choice is $\chi^a=(1,0,\ldots, 0)$ up to normalization, with
$C$ the round $S^{n-1}$ sphere with volume element $d\Theta$, so that
\begin{equation} 
\mu(t) = -\frac{1}{8\pi (n-2)} \int   E_{00} d\Theta \, .  \label{AM2}
\end{equation}

Now, up to a scale factor, on $\scri$ the ``spray'' of future directed
null geodesics at a point is parameterised by an $n-2$ dimensional
sphere.  Thus, for example, in $n+1=4$ dimensions there is a circle
worth of null directions on $\scri$ at every point, while in $n+1=5$
dimensions there is a 2-sphere worth of null directions. 
Taking $d\Xi$ as the
volume element on the round $S^{n-2}$, we then take the average
\begin{equation}\label{massaspect}
- \int E_{ab} \, \gamma^a\gamma^b \, \,  d\Xi =
-\frac{n }{n-1} \, \,{\rm vol}(S^{n-2}) \, E_{00}. \label{av1}  
\end{equation} 
Here we have used (a) the fact that the integral is performed over the
future null directions {\it at a point}, so that $E_{a b}$ is in fact
independent of these directions and (b) the fact that $E_{ab}$ is
trace-free.  Averaging the time-delay (\ref{TD1})
over all possible null directions as in (\ref{massaspect}) and over
all points on the boundary $S^{n-1}$, we get
\begin{equation}
\int \Delta T \, \, d\Xi d\Theta = \frac{8 \pi \,  {\rm
vol}(S^{n-2})}{(n-1)}\, \,\om^n \,  \int_{t_0}^{t_0+\pi}
\sin^n(t-t_0) \,\, \mu(t) \, \,d t + \cO(\om^{n+1}). 
\end{equation}    
Since the time delay $\Delta T \geq 0 $, this implies positivity of  the
weighted time-averaged mass, 
\begin{equation} 
\label{massav}
\mw = \int_{t_0}^{t_0 +\pi}  \sin^n(t-t_0)\,\, \mu(t) \,\,dt \geq 0.
\end{equation} 
 If the matter flux $\int_V T_{ab}
\chi^a \chi^b dV$ vanishes on $\scri$ for the patch $V \subset \scri$
bounded by spacelike cuts of $\scri$ at $t_0$ and $t_0+\pi$, then
$\mu(t)$ is constant in $V$ \cite{am,ashdas}, so that 
the instantaneous mass $\mu(t) \geq 0$ in $V$.
We have thus proved the following:

\noindent {\bf Theorem:} Assume that an $n+1$ dimensional
asymptotically AdS spacetime $(M, g)$ has a holographic dual on its
conformal boundary $\scri$. Then the requirement of holographic
causality implies that the weighted time-averaged Ashtekar-Magnon mass
$\mw$ for the spacetime is non-negative for all $t_0 \in \re$. In particular,
when the matter flux at $\scri$ is zero for a time period of
at least $\pi$, the instantaneous Ashtekar-Magnon mass is non-negative
over that same time period.

\noindent {\sl Remark (i):} In \cite{eric} the weight in $\mw$ was
assumed to be uniform, but we have shown here that it is indeed
non-uniform.  If this weight had been a delta function instead of
$\sin^n (t-t_0)$, one could conclude that the instantaneous mass is
non-negative in general. However, $\sin^n(t-t_0)$ only approaches a delta
function as $n \to \infty$, so for finite $n$ the weight has
non-vanishing 
width. While this appears to be a drawback of the focusing method
employed here, one may be able to extract some useful information from
it. In particular, (a) $\mu(t)$ cannot be negative for a time scale of
$\sim \pi$, and (b) since $\mw \geq 0 \, $ for all $t_0$, if $\mu(t)<0$
on time-scales $\tau < \pi$, then it must be compensated for by a
positive contribution from $\mu(t)$ for a time $\pi -\tau$ with
``interest''. What does this say about classical stability? One might
suppose that if the mass of the spacetime becomes negative at any time
that this would lead to an increasingly negative mass. However, the
fact that $\mw \geq 0$ {\it for all} $t_0$ clearly prevents such an
instability.  \\
\noindent {\sl Remark (ii):} The elements of our proof do not suffice
to prove the converse of the theorem, namely that positivity of mass
is sufficient to ensure holographic causality.  Positivity of mass
does not imply that the function $-E_{ab} \gamma^a \gamma^b$ is
non-negative everywhere, and by choosing $\gamma$ and $\gamma'$
suitably, the time delay could be made negative. Instead of assuming
positivity of mass however, if we were to assume null genericity, null
geodesic completeness and a null energy condition, then a focusing
theorem \cite{he,tipler,borde} can be used to show that all bulk
geodesics  have non-negative time-delay with respect to boundary null
geodesics\footnote{The focusing theorem was used to complete the proof of
positivity of energy in conjunction with the time delay formula in
\cite{eric}.}.  This theorem states that in such spacetimes any
complete causal geodesic must contain conjugate points. The existence
of conjugate points means that the null geodesic focuses, so that the
fastest null geodesics always lie on the boundary.  \\
\noindent {\sl Remark (iii):} In \cite{kgt}, it was claimed that any
asymptotically AdS spacetime must satisfy an integrated weak energy
condition if it is to have a causal holographic boundary dual. While
the focusing theorem {\it implies} positive time delay, it is not
clear how the converse conclusion of \cite{kgt} may be reached. \\
\noindent {\sl Remark (iv):} Finally, we briefly comment on the
relation to the counterterm mass definition of \cite{balkraus}, where
the mass $\mu_{ct}$ is calculated by adding certain local counterterms
to the action. In 4-dimensions it was shown in \cite{ashdas} that this
mass definition is identical to the AM mass, so that holographic 
causality implies the positivity of $\mu_{ct}$ by our theorem. In 5
dimensions, however, these masses explicitly do not match, and their
difference $\mu_{ct} - \mu$ can be interpreted as a non-zero Casimir
energy. Using our method, however, holographic  causality cannot be used to
predict the sign of this Casimir energy term.

Now, the time delay formula (\ref{TD1}) is valid only for
asymptotically AdS spacetimes with $\scri$ of topology $S^{n-1}\times
\re$.  However, asymptotically locally AdS spacetimes with
non-spherical boundary are also candidates for a holographic theory
\cite{db,hm,ejm,ssw}, and several examples of spacetimes with such
boundaries are known \cite{topbhs}. Positive energy theorems have not
been proved for such spacetimes --- indeed, the AdS soliton \cite{hm},
which has a Ricci flat $\scri$, has negative AM mass\footnote{The soliton is protected from a potential instability due to 
a unique scaling {\it isometry} between solitons of different mass
parameters \cite{gsw1}.}.  Motivated by the AdS/CFT 
correspondence, Horowitz and Myers \cite{hm} proposed a new positive
energy conjecture which states that the AdS soliton is the unique
ground state in its asymptotic class. It is therefore of interest to
ask how the considerations in this current work extend to such
spacetimes.

We now show by explicit calculation that despite having negative mass,
the AdS soliton has a strictly positive time delay when $n \geq 3$, so
that the boundary theory is causal in this restricted sense. The $n+1$
dimensional AdS soliton has metric
\begin{equation}
\label{soliton}
ds^2 = -r^2 dt^2 + \frac{1}{V(r)} dr^2 + V(r) d\phi^2 + r^2
\sum\limits_{i=1}^{n-2}(d\theta^i)^2 . \label{Intro1}
\end{equation}
Here $V(r)=\frac{r^2}{l^2}(1 - \frac{r_0^n}{r^n})$, with
$\ell^2=-\frac{n(n-1)}{2\Lambda}$, $r_0$ a constant and $n> 2$. The
solution is non-singular provided $\phi$ is periodic with period
$\beta_0=\frac{4\pi\ell^2}{n \, r_0}$. 
Putting $x\equiv 1/r$,  (\ref{soliton}) can be re-expressed as 
\begin{equation}
ds^2 = \frac{1}{x^2}(-dt^2 + ds_n^2),  
\end{equation} 
where 
\begin{equation}
ds_n^2 = \frac{\ell^2 }{(1-r_0^nx^n)} dx^2 + \frac{(1-r_0^nx^n)
}{\ell^2} d\phi^2+ \sum_i^{n-2} d\theta_i^2, 
\end{equation} 
is the spatial part of the optical metric $d\ws^2 = x^{2}ds^2$ . Thus,
the time taken by a null signal is $\Delta t = \int ds_n.$ 

Now, consider a null geodesic from a source S on $\scri$, which has
angular momenta $(\vec L, J)$, where $\vec L \equiv (\dot \theta_1,
\cdots , \dot{\theta}_{n-2})$, $J\equiv (1-r_0^nx^n)\dot \phi$ and
$\dot{\vec L}=\dot J=0$ along any geodesic where $\dot A \equiv dA/ds_n$.
The time taken for this geodesic to get back to $\scri$ via the bulk
(and {\it not} along the boundary) is then
\begin{equation}
\Delta t' =  \sqrt{ |\Delta \vec \Theta|^2
+\biggl(\frac{4\ell^2}{r_0^2}\biggr) \, \, 
{(1-k^2)^{\frac{2-n}{2n}} \,\, I(n)^2}}, \label{TD3}    
\end{equation} 
where $|\Delta \vec \Theta|$ is the optical distance traversed in the $\vec
\Theta\equiv (\theta_1, \cdots , \theta_{n-2})$ directions, $k^2=
\ell^{\, -2} \, J^2 \,  (1-L^2)^{-1}$ and 
\begin{equation} 
I(n)=\int_0^1 \frac{1}{\sqrt{1-X^n}} \, dX = \sqrt{\pi}\,\,
\frac{\Gamma(\frac{n+1}{n})}{\Gamma(\frac{n+2}{2n})}. 
\end{equation} 
If the bulk null geodesic is received by some observer R on $\scri$,
the shortest time taken by a boundary null geodesic to get from S to R
satisfies the inequality,
\begin{equation}
\Delta t \leq \sqrt{|\Delta \vec \Theta|^2 + \biggl( \frac{4
\ell^2}{r_0^2}\biggr) \frac{\pi^2}{n^2}}.
\label{TD2}    
\end{equation}
If the fastest curve, necessarily a null geodesic, were to lie in the
bulk, then $\Delta t' < \Delta t$ for that geodesic. But we see that
this cannot be the case. Comparing the times $\Delta t$ and $\Delta t'$ we see
that the time delay $\Delta T=\Delta t'-\Delta t$ is positive whenever
\begin{equation} \label{ineq}
\frac{n}{\pi}\,  I(n) \, \,  (1-k^2)^{\frac{2-n}{4n}} \geq 1. 
\end{equation} 
Now, $(1-k^2)^{\frac{2-n}{4n}} \geq 1$ for $n\geq 2$, the equality
being satisfied only when $n=2$ and/or $J=0$. Moreover, one can easily
show that $\frac{n}{\pi}I(n) \geq 1$ for $n\geq 2 $ --- indeed, it
grows approximately linearly with $n$.  Thus, the time-delay is zero
only for $n=2$. This is not surprising, since in 3 spacetime
dimensions, the AdS soliton is in fact global AdS.  It is also
satisfying to note that the existence of the time delay is in fact
independent of the apparent mass ``parameter'' $r_0$. This is
consistent with the fact that solitons of different $r_0$ are in fact
isometric to each other via a scaling isometry \cite{gsw1}.

Thus, the AdS soliton, despite having negative mass, satisfies
holographic causality. However, in light of the Horowitz-Myers
conjecture, it is interesting that there is no bulk null geodesic
which realizes the bound set by causality or alternatively, by the
focusing theorem.  What this means for the conjecture is a question
under current investigation \cite{es}.

\noindent {\bf Acknowledgments:} This work was partially supported by
grants from the Natural Sciences and Engineering Research Council
(Canada). SS was supported by a post-doctoral fellowship from the
Pacific Institute for the Mathematical Sciences.

\end{document}